\documentstyle[12pt,fleqn,twoside]{article}

\def\pr{{ Phys. Rev.}\ }

\def\prl{{ Phys. Rev. Lett.}\ }
\def\pl{{ Phys. Lett.}\ }

\def\nat{{ Nature}\ }
\def\apj{{ ApJ}\ }

\def\aaa{{A\&A}\ }

\def\mnras{{MNRAS}\ }

\def\etal{{et al.}\ }

\newcommand{\mincir}{\raise -2.truept\hbox{\rlap{\hbox{$\sim$}}\raise5.truept
\hbox{$<$}\ }}
\newcommand{\magcir}{\raise -2.truept\hbox{\rlap{\hbox{$\sim$}}\raise5.truept
\hbox{$>$}\ }}
\newcommand{\minmag}{\raise-2.truept\hbox{\rlap{\hbox{$<$}}\raise 6.truept\hbox
{$>$}\ }}
\newcommand{\bx}{\mbox{\bf x}}
\newcommand{\by}{\mbox{\bf y}}
\def\ba{\begin{eqnarray}}
\def\ea{\end{eqnarray}}
\def\be{\begin{equation}}
\def\ee{\end{equation}}

\def\hm{h^{-1}{\rm Mpc}}

\def\etal{{et al.}}

\let\gam=\gamma

\title{ {\bf Large-Scale Clustering in Bubble Models}}
\author{
{\bf Luca Amendola}$^1$ \& {\bf Stefano Borgani}$^2$ \\ ~\\
{\em $^1$Osservatorio Astronomico di Roma}\\
{\em Viale del Parco Mellini, 84}\\
{\em I--00136 Rome, Italy} \\ ~\\
{\em $^2$INFN, Sezione di Perugia,}\\
{\em c/o Dipartimento di Fisica dell'Universit\`a,} \\
{\em via A. Pascoli, I--06100 Perugia, Italy} }

\date{}
\begin{document}
\maketitle
\vspace{0.5truecm}
\centerline{\sl Monthly Notices of the Royal astronomical Society, submitted}
\thispagestyle{empty}

\setcounter{page}{0}
\newpage
\centerline{\large \bf Large-Scale Clustering in Bubble Models}
\vspace{.3in}
\centerline{Luca Amendola$^1$ \& Stefano Borgani$^2$}
\vspace{.2in}
\centerline{$^1${\it Osservatorio Astronomico di Roma}}
\centerline{\it Viale del Parco Mellini, 84}
\centerline{\it I-00136 Rome, Italy}
\vspace{.1in}
\centerline{$^2${\it INFN - Sezione di Perugia}}
\centerline{\it c/o Dipartimento di Fisica dell'Universit\`a}
\centerline{\it Via A. Pascoli, I-06100 Perugia, Italy}
\vspace{.2in}
\centerline{ABSTRACT}
\baselineskip 14pt
\begin{quote}
We analyze the statistical properties of bubble models for the large-scale
distribution of galaxies. To this aim, we realize static simulations, in
which galaxies are mostly randomly arranged in the regions surrounding
bubbles. As a first test, we realize simulations of the Lick map, by
suitably projecting the three-dimensional simulations. In this way, we are
able to safely compare the angular correlation function implied by a bubbly
geometry to that of the APM sample. Quite remarkably, we find that several
bubble models provide an adequate amount of large-scale correlation, which
nicely fits that of APM galaxies. Further, we apply the statistics of the
count-in-cell moments to the three-dimensional distribution and compare
them with available observational data on variance, skewness and kurtosis.
Based on our purely geometrical constructions, we find that a well defined
hierarchical scaling of higher order moments up to scales $\sim 70\hm$. We
show that this must be expected for any non-Gaussian distribution in the
weak-coherence regime. The overall emerging picture is that the bubbly
geometry is well suited to reproduce several aspects of large-scale
clustering. Furthermore, the statistical test we apply are able to
discriminate between different models. We find that models
with fixed bubble radius have troubles to account for the angular
correlation of APM galaxies, while a shallow spectrum fails to reproduce
the observed three-dimensional skewness. A model with a rather steep
spectrum of bubble of radii is rather adequate to account for the
observational tests that we considered.
\end{quote}
\vspace {.1in}
\normalsize
\newpage

\section{\normalsize\bf Introduction}
The most striking feature of the large-scale structure of the Universe is
the remarkable variety of coherent structures and the extension of the
involved scales. Its interpretation really represents a difficult task for
any theoretical model about the origin of primordial fluctuations. The most
eminent victim of such a largely structured Universe is probably the
standard cold dark matter (CDM) scenario. In its minimal version, this
model assumes initial adiabatic and Gaussian fluctuations, dominated by
highly non-relativistic non-baryonic particles in a $\Omega=1$ Universe,
with galaxies substantially more clustered than the underlying dark matter
(see, e.g., Blumenthal et al. 1984; see also Davis {  et al.}
1992 and Liddle \& Lyth 1993, for recent reviews about CDM). Although
successful at small and intermediate scales, this model has been shown to
have serious troubles in accounting for clustering data at scales $>20\hm$,
such as clustering of rich clusters ({  e.g.}, White {  et al.} 1987),
large-scale galaxy angular correlation (Maddox {  et al.} 1990; Collins
{  et al.} 1991), large-scale galaxy peculiar motions (Vittorio,
Juszkiewicz \& Davis 1986; Bertschinger {  et al.} 1990), count-in-cell
statistics of IRAS-QDOT (Saunders {  et al.} 1991) and Stromlo-APM
(Loveday {  et al.} 1992) galaxies, large-scale power spectra
(Peacock 1991; Feldman, Kaiser \& Peacock 1993), abundance
of rich clusters (White, Efstathiou \& Frenk 1993).

Several adjustments of the standard CDM model have been proposed in order
to overcome its deficit of large-scale power, which
can be divided into two main categories; those which modify the scheme for
the production of primordial inhomogeneities and those which try to suitably
modify the prescription for galaxy identification out of dark matter
fluctuations. The first category includes models with primordial
power-spectrum tilted away from the Zel'dovich one, predicted by
standard inflation ({  e.g.}, Lucchin \& Matarrese 1985; Cen \etal
1992), low-density Universe with non-vanishing cosmological constant
($\Omega_{matter} \simeq 0.2,~\Omega_{matter}+\Omega_{\Lambda}
 =1$; {  e.g.}, Efstathiou, Sutherland \& Maddox
1990), non-Gaussian initial fluctuations (Moscardini {  et al.} 1991;
Weinberg \& Cole 1992), inclusion of a hot particle component supplying
large-scale power (Valdarnini \& Bonometto 1985; Achilli, Occhionero \&
Scaramella 1985; Klypin {  et al.} 1992). The second class of
non-standard CDM models includes low-bias mechanisms of galaxy formation,
in which the present time corresponds to a dynamically more evolved stage
(Couchman \& Carlberg 1992; see, however, Moscardini {  et al.} 1993), and
``cooperative" non-local biassing (Bower {  et al.} 1993), in which galaxy
formation is triggered by astrophysical mechanisms, able to introduce
large-scale coherence ({  e.g.}, inhibition of galaxy formation due to
high-redshift QSO reionization; Babul \& White 1991).

In this paper we follow a purely geometrical approach by modelling the
large-scale clustering by means of a distribution of spherical bubbles,
with galaxies mostly arranged in the volume not occupied by such bubbles,
while leaving them almost devoid. Therefore, we are not concerned with the
dynamics underlying the above models for large-scale structure formation.
Instead, we attempt to parametrize their dynamical content through a
suitable geometrical construction.

Bubble models have been already considered in several contexts.
Generation of bubbles can be provided by invoking primordial
phase-transitions, occurred along with the inflationary expansion
({  e.g.}, La 1991). The only theory to date for primeval bubble
generation, which also predicts a definite spectrum of bubble sizes, is the
extended inflation (La \& Steinhardt 1989;
for a review see Kolb 1991). In this scenario, bubbles are
generated as true-vacuum regions, tunnelling out from the initial
false-vacuum state and subsequently stretched to cosmological sizes by the
accelerated expansion. In the original model, the number of bubbles $N_B(R)$
with radius larger than $R$ is roughly a power-law,
\be
N_B(R)~=~(R_M/R)^p\,,
\label{eq:rsp}
\ee
where $p$ is related to the microphysical parameters of the phase transition
and $R_M$ is the maximum
bubble radius inside the horizon, also fixed by the model.
If the
inflation is driven by the Brans-Dicke field
coupled through the constant $b$, like in La \& Steinhardt (1989),
then $p=3+4/(b+1/2)$. The determination of the slope $p$ of
the present bubble spectrum may then be related to the fundamental
constant $b$. A potential
problem in this mechanism of bubble production lies in the predicted $R_M$
values, which, in all the models considered
so far, are either in conflict with the microwave background isotropy
or too small to be of cosmological interest (Liddle \& Wands 1991). A
primordial generation of bubbles clearly fixes non-Gaussian initial
conditions, since the expectation of finding underdense regions is larger
than that of finding overdense regions, and the resulting probability
density function of the density field is negatively skewed.

A bubbly galaxy distribution naturally arises also in the explosion scenario
(Ostriker \& Cowie 1981; Ikeuchi 1981). Differently from the gravitational
instability picture, in this scenario energy perturbations of
non-gravitational origin drive material away from the seeds of the
explosions, sweeping primordial gas into dense, expanding shells. As these
shells cool, their fragmentation could give a further generation of objects,
which again explode, thus amplifying the process and giving rise to
large-scale structure formation. At the end, galaxies are arranged on spherical
shells and rich clusters are expected to be placed at the intersection of
three expanding bubbles. A variety of physical mechanisms may generate such
explosions, such as supermassive stars or supernovae from the earliest
galaxies. However, it is at present not clear whether such energy
sources are sufficient to create the large voids that are observed.
Moreover, suitable initial conditions are anyway needed to generate
primordial objects, which act as seeds of explosions.
It is also to be remarked that the recent analysis of the COBE FIRAS
spectrum (Wright \etal 1993) seems to rule out the explosions scenario,
since the energy release would have distorted the cosmic blackbody
spectrum above the observational bounds.

A more conventional mechanism for nearly spherical bubble generation is
provided by gravitational instability in the mildly non-linear regime. In
the pancake scenario, overdense regions tend to increase their asphericity
during the gravitational collapse, while underdense regions expand and
become more and more spherical. As a result, the most part of space turns
out to be occupied by almost devoid spherical regions, with galaxies
arranged in sheet-like structures around them. It is however clear that in
order for gravitational dynamics to produce as large voids as observed, a
primordial fluctuation spectrum is required, which produces an adequate
amount of large-scale fluctuation power, probably more than supplied by
standard CDM.

Purely geometrical random bubble distributions have been already shown to
be able to account for some features of large-scale clustering. In the
framework of the explosion model, Weinberg, Ostriker \& Dekel (1989)
used a random distribution of expanding shells with both
a power-law and a uniform distribution
of radii. After identifying clusters at the intersection of three of such
shells, the statistics of their observed distribution is quite well
reproduced. A detailed description of the galaxy distribution in terms of
Voronoi tessellation (van de Weygaert \& Icke 1989;
Yoshioka \& Ikeuchi 1989; van de Weygaert 1993)
also reveals that a topology dominated by big voids surrounded by a
sheet-like galaxy distribution gives a fair representation of several
clustering features.

In this paper, we analyze statistical properties for simulations of
bubbly galaxy distributions and compare them with similar tests, already
presented in the literature, about the observed galaxy clustering. This
allows us to check in detail the ability of this purely geometrical model
to account for the large-scale texture of the Universe. In the simulations, the
centers of spherical bubbles are randomly distributed and the radii are
chosen either fixed or taken from a power law spectrum, like that of
Eq.(\ref{eq:rsp}), with a lower cut-off radius $R_m$ of the order of
a few megaparsecs.
 A list of the parameters involved in each of the models that we
will consider is presented in Table I. A letter F marks the models with fixed
radius size. The model parameters are always
tuned in such a way that the resulting bubble volume is 90\% of the
total volume. Galaxies are randomly arranged in the space not occupied by
bubbles, while an unclustered component, a fraction $f_u$ of the total number
of particles, is also added to account for a
residual galaxy population inside the underdense regions. The simulation
box boundary is taken to be periodic. By varying the
model parameters, we generate simulated galaxy distributions having
different amounts of large-scale clustering. It is clear that using such
simple simulations we completely neglect the presence of gravitational
clustering at the scales of non-linearity. For this reason, we restrict our
analysis to larger scales ($>5\hm$), where non-linear gravitational
dynamics does not work and the clustering should be accounted for just by
the large-scale geometry. The remarkable advantage of adopting such
non-dynamical simulations
is that we can reach very large scales (we take $450\hm$ for the
simulation box side), with a negligible computational cost,
even when we average the results over several realizations. In a previous
work (Amendola \& Occhionero 1993), the bubble model has
been compared to some observational constraints,
such as the large-scale shape of $1+\xi(r)$ (here, $\xi(r)$ is the galaxy
spatial 2-point correlation function) and the upper limits on anisotropy
and spectral distortion of the cosmic microwave background. In particular,
it was found that power-law bubble models can fit the observed
spatial correlation function if
\be\label{fits}
R_M \approx (p/130)^{-1.3}
\ee
(tested in the range $p=4-11$)
and if about 90\% of the total volume is occupied by the bubbles.
The latter requirement fixes the cut-off size $R_m$, i.e. the radius
of the smallest bubbles. All the power-law models we test in this
paper roughly obey the prescriptions above.
If bubbles were generated
before decoupling, Amendola \& Occhionero (1993) also show that power
law spectra pass the cosmic microwave background tests only for
$6<p<11$; roughly speaking,
shallower spectra have too many large bubbles,
while steeper spectra too many small bubbles.

In Figure 1 we
plot the resulting galaxy distribution for different bubble models, having
both fixed and variable radii, according to the power-law spectrum of
Eq.(\ref{eq:rsp}). Each panel reports the projection in the $x-y$ plane of
galaxies in a 15$\hm$ slice in the $z$ direction. It is apparent the
remarkable variety of structures generated by the bubble geometry, with
rich clumps, filaments and underdense regions having sizes of some tens of
Mpcs.
Also the effect of taking different model parameters is visible in this
plot. Since each realization is generated with the same phase assignment,
the persistence of several structures can be directly compared along the
sequence of the model parameters. Fixed radius and steep spectra produces
more isolated and concentrated galaxy aggregates, while shallower spectra
turns into the presence of larger bubbles and, consequently, of more
extended coherent structures. These differences will be quantified in more
details in the remainder of this paper.

The plan of the paper is as follows.

After generating artificial Lick maps by projection of the spatial
simulations, we analyze in Section 2 the resulting angular correlation
functions and compare them with that observed for the APM sample (Maddox
{  et al.} 1990). Section 3 is devoted to the analysis of moments of
cells counts (variance, skewness and kurtosis). The comparison with
analogous results for observational data sets allows us to further restrict
the parameter space for the permitted bubble models and to assess the
reliability of this geometrical interpretation of large-scale clustering.
In Section 4 we discuss our results and draw the main conclusions.

\section{\normalsize\bf Angular correlation}
As a first test of large-scale clustering, we investigate the projected
(angular) correlation function implied by the bubble geometry. In order to
implement this analysis, we generate artificial Lick galaxy catalogues,
following the same procedure adopted by Coles {  et al.} (1993, hereafter
CMPLMM) and Moscardini {  et al.} (1993). The observer is placed at the
center of the simulation box. Then, the original box is reflected, so to
generate a three dimensional observational cone having width and depth
adequate to reproduce the Lick map. Although the characteristic depth of
the Lick map is $\sim 210\,h^{-1}$ Mpc, also very luminous galaxies at a
much larger distance are included. Taking for the luminosity function the
Schechter-like expression
\be
\Phi(L) \propto \left({L\over L^*}\right)^{-\alpha}\,
\exp\left(-{L\over L^*}\right)
\label{eq:phil}
\ee
with the parameters given by Efstathiou, Ellis \& Peterson (1988; $\alpha=
-1.07$ and $M^*=-19.68$ for the absolute magnitude associated to the
characteristic luminosity $L^*$), few galaxies up to $\sim 700\,h^{-1}$ Mpc
are included. Taking a simulation cube of $450\,h^{-1}$ Mpc aside, we
select galaxies up to the distance of $675 \hm$. With this kind of
construction the observational cone encompasses the range of galactic
latitude $|b^{\mbox{\tiny II}}|\ge 45^\circ$. The procedure to assign
luminosities to the
galaxies is the same as described by CMPLMM. We suitably truncate $\Phi(L)$
at the faint and bright tails, assign the apparent magnitude to a galaxy at
a given distance by including $K$-correction and curvature effects.
Therefore, we project only those galaxies brighter than the limiting
apparent magnitude of the Lick map ($m \le 18.8$). We tune the mean galaxy
separation in the simulation box so to end up with a total number of
projected galaxies of the same order of that included in the Lick map for
$|b^{\mbox{\tiny II}}|\ge 45^ \circ$ ($\sim 316,000$). As observed by CMPLMM,
although the
observational cone includes galaxies belonging to different replica of the
same simulation box, no substantial spatial periodicity should affect the
projected clustering pattern, since each galaxy has a very small
probability to be selected more than once. This problem is even less
important in our case, being our simulation cube much larger than that used
by CMPLMM, which used three levels of replicated boxes. After generating
the projected galaxy distribution, we collect them in counts with $10\times
10$ arcmin cells, so to reproduce the observational setup of the Lick map.

Simulating angular galaxy samples, like the Lick map, has been proved to be
a reliable approach to compare data and simulations. In fact, the lack of
redshift information in angular analysis is largely compensated by the
extension of the available data sets. CMPLMM generated artificial Lick maps
from N-body simulations, based on assuming both Gaussian and skewed CDM
initial conditions, in order to investigate the topology of the
two-dimensional galaxy distribution. Moscardini {  et al.} (1993) used
the same approach to investigate the angular correlation functions of such
simulated samples. Borgani {  et al.} (1993) extracted synthetic samples
of galaxy clusters from the same Lick map simulations and applied several
statistical tests to compare them with clusters selected from the real Lick
map (Plionis, Barrow \& Frenk 1991).

Here, we analyze the angular correlation function, $w(\vartheta)$, of our
Lick simulations and compare them with the APM data, as provided by Maddox
{  et al.} (1990). The adopted estimator of the angular two-point
correlation function is
\be
w(\vartheta)~=~{\left< n_in_j\right> _\vartheta
\over \left< {1\over 2}(n_i+n_j)\right> _\vartheta^2} -1\,.
\label{eq:wth}
\ee
Here, $n_i$ and $n_j$ are the galaxy counts in the $i$--th and $j$--th
cell, respectively, placed at separation $\vartheta$. The average is taken
over all cell pairs with separation $\vartheta$. Since our simulations of a
bubbly Universe contain only geometry, we are not interested in the
small-scale behaviour of $w(\vartheta)$, which is expected to be determined
by non-linear gravitational dynamics. Considering only separation above
$1^\circ \!\!.5$, which corresponds to $\simeq 5\,h^{-1}$ Mpc at the Lick
depth (Maddox \etal 1990), allows us to considerably reduce the
computational time. We group the small cells to form counts in $1^\circ
\times 1^\circ$ cells, and $w(\vartheta)$ is evaluated according to
eq.(\ref{eq:wth}), being $n_i$ and $n_j$ the counts in such larger cells.

The results of our analysis are shown in Figure 2, where we plot the angular
correlation function for a list of bubble models, as compared to the APM
data, suitably rescaled to the Lick depth. In Table I we report the
relevant parameters for the models we are considering. The upper left panel
is for the fixed-radius model. It is apparent that the model with
$R_f=12\,h^{-1}$ Mpc displays a deficit of correlation strength at
separations $\ge 3^\circ$, while providing the correct small scale
amplitude. Increasing the radius at $R_f=14\,h^{-1}$ Mpc increases
$w(\vartheta)$, thus providing an excess of small-scale clustering. The
upper right panel is for the variable radius models, with power index $p=8$
and for two different values of the maximum bubble radius. It is apparent
that allowing for a variation of the bubble radius sensibly improves the
agreement. The shape of $w(\vartheta)$ is now quite well reproduced at all
the relevant angular scales. In the lower left panel we check the effect of
taking an even wider spectrum of bubble radii, by lowering the $p$ value.
Also in this case, the overall shape of $w(\vartheta)$ is remarkably well
reproduced. Finally, in the lower right panel we verify the effect of
taking different amounts of unclustered galaxies within a fixed model. The
plot shows that the $w(\vartheta)$ shape is left substantially unchanged
apart from a variation of the small scale-amplitude. Therefore, increasing the
percentage of unclustered galaxies amounts essentially to shift
$w(\vartheta)$ downwards. This suggests that, once a model is shown to produce
the
correct correlation shape, its normalization can be fixed a posteriori by
choosing a suitable amount of random galaxies. It is however clear that
this procedure is not completely arbitrary, since observational constraints
on the percentage of weakly clustered galaxies in voids suggests that they
should be $\magcir 10\%$ of the whole population. Note that several models
produces a $w(\vartheta)$ slope at the smallest considered scales, which
remarkably matches that for the APM sample. Since our simulations do not
include gravitational dynamics, which is expected to be responsible for the
small-scale $w(\vartheta)$ profile, this may suggest that the bubbly geometry
can be rather adequate also to account for mildly non-linear clustering.

The above results about angular correlation analysis indicate that the
geometrical pattern provided by a random bubble distribution is able to
provide a rather adequate amount of the large-scale power traced by the
largest available projected galaxy samples. However, although several
models are clearly ruled out, and others, such as that with fixed radii,
are only marginally acceptable, the $w(\vartheta)$ analysis seems not to be
able to pick out one preferred model. For instance, looking at the plots of
Figure 2, it is not clear whether a steep spectrum of bubble radii (with
$p=8$) with rather small $R_M$ is better than a having $p\simeq 5$ and a
larger $R_M$ value. In order to further discriminate between different
models, we apply in the following the higher-order statistics of
count-in-cell moments to the three-dimensional distributions.

\section{\normalsize\bf Moments of cell-counts}
\subsection*{\normalsize 3.1 ~~The method}
The analysis of moments of count in cells has been recently widely employed
to analyze the observed galaxy distribution and it has been found to be
a powerful statistical test of large-scale clustering (Efstathiou \etal
1990; Saunders \etal 1991; Bouchet, Davis \& Strauss 1992; Coles \& Frenk 1991;
Gatza\~naga 1992; Loveday \etal 1992). The main advantage of this kind of
analysis lies in the fact that it is relatively easy to implement and turns
out to be particularly suited to investigate the statistics at large
scales, where statistical noise severely limits the usage of correlation
functions.

In order to characterize the statistics of the moments of counts,
let us introduce $\delta _R(\bx)=\rho_R(\bx)/\bar \rho -1$ as the relative
fluctuations of the
density field $\rho_R(\bx)$, smoothed over a suitable angular scale $R$ by
a window function $W_R(\bx)$. Therefore,
\be\delta_R(\bx)~=~{1\over V_R}\,\int d^3y \delta(\by)\,W_R(\bx-\by)\,,
\label{eq:dr}
\ee
being $V_R=\int d^3x \,W_R(\bx)$ the volume
encompassed by the chosen window.
At the scale $R$, the statistics of the fluctuation field can
be described by using the moments $\mu_n(R)\equiv \left<\delta_R^n\right>$.
Let us consider $\delta_R$ as a random variable with zero mean and let
$P(\delta_R)$ be its probability density function. Accordingly, its
moment of order $n$ is
\be
\mu_n(R)~=~\int d\delta_R \,P(\delta_R)\,\delta_R^n\,.
\label{eq:mun}
\ee
The corresponding generating function
\be
M(\phi)~=~\left<e^{\phi \delta_R} \right>
{}~=~\int d\delta_R\,P(\delta_R)\,e^{\phi \delta_R}
\label{eq:mphi}
\ee
is defined in such a way that the $\mu_n$ moments represent the
coefficients of its McLaurin expansion,
\be
M(\phi)~=~\sum_{n=0}^\infty{\mu_n \over n!}\,\phi^n~~~~~~~~~~~;~~~~~~~~~~~
\mu_n~=~{d^nM(\phi)\over d\phi^n}\bigg|_{\phi =0}\,.
\label{eq:mun1}
\ee
An equivalent statistical description can also be given in terms of the
{\em cumulant} generating function
\be
K(\phi)~=~\log{M(\phi)}~=~\sum_{n=0}^\infty{\kappa_n \over n!}\,\phi^n
{}~~~~~~~~~~~;~~~~~~~~~~~
\kappa_n~=~{d^nK(\phi)\over d\phi^n}\bigg|_{\phi =0}\,.
\label{eq:kphi}
\ee
The cumulants $\kappa_n$ can be related to the moments $\mu_n$ by
subsequently differentiating the corresponding generating functions. At the
lowest orders, it is
\be
\kappa_2~=~\mu_2~~~~~;~~~~~\kappa_3~=~\mu_3~~~~~;~~~~~
\kappa_4~=~\mu_4-3\mu_2^2\,.
\label{eq:kmu}
\ee
with more cumbersome relations holding at higher orders. In turn, the
cumulant of order $n$ represents the average value of the connected
$n$-point correlation function $\xi_n$ inside the window volume $V_R$:
\be
\kappa_n(R)~=~{1\over (V_R)^n}\,\int_{V_R}
\left\{\prod_{i=1}^nd^3x_i\,W_R(\bx_i)\right\}\,\xi_n(\bx_1,...,\bx_n)\,.
\label{eq:mnxi}
\ee
Under the assumption that the 2-point function behaves like a pure
power-law, $\xi(r)=(r_o/r)^\gamma$, at the second order the variance is
$\sigma^2=\kappa_2=J_2\,\xi_2$, where
\be
J_2~=~{1\over (V_1)^2}\, \int_{V_1} d^3x_1\,\int_{V_1} d^3x_2\,
|\bx_1-\bx_2|^{-\gamma}W_1(\bx_1)\,W_1(\bx_2)
\label{eq:j2}
\ee
arises after changing the variables in the integrals of eq.(\ref{eq:mnxi}),
according to $\bx_i \to \bx_i/R$. In this way, $V_1$ represents the volume
encompassed by the window of unit radius, $W_1(\bx)$, so that $J_2$
is a dimensionless coefficient, which only depends on the window profile
and on the slope $\gamma$ of the 2-point function.

At higher orders, a popular model to describe the scaling of the
moments is represented by the hierarchical expression
\be
\kappa_n~=~S_n\,\kappa_2^{\eta_n}
\label{eq:hie}
\ee
where $\eta_n=n-1$ and $S_n$ are suitable coefficients, which depends on the
moment order, other than on the shape of the window function. For instance,
the hierarchical 3-point function reads $\xi_3=3Q \xi_2^2$,
so that eq.(\ref{eq:mnxi}) gives $\kappa_3=3Q(J_3/J_2^2)\xi_2^2$,
where
\be
J_3~=~{1\over (V_1)^3} \int_{V_1}
\left\{\prod_{i=1}^3 d^3x_i\,W_1(\bx_i)\right\}\,
|\bx_1-\bx_2|^{-\gamma}|\bx_1-\bx_3|^{-\gamma}|\bx_2-\bx_3|^{-\gamma}
\label{eq:j3}
\ee
plays the same role as $J_2$ for $n=3$. On a theoretical ground the
hierarchical recurrence relation is expected to be dynamically generated by
strongly non-linear gravitational clustering ({  e.g.}, Peebles 1980). In
this regime, the closure of the BBGKY equations provides hierarchical
correlations with specified $S_n$ coefficients (Fry 1984a; Hamilton 1988).
Also in the mildly non-linear regime, second-order perturbative approaches
to the evolution of gravitational clustering generates a hierarchical
sequence of correlations ({  e.g.}, Fry 1984b; Juskiewicz \& Bouchet
1992), but with values of the coefficients $S_n$  which are in general
different from those arising from non-linear gravitational clustering
(Lahav \etal 1993).

Suppose now to represent the continuous density field with a discrete point
distribution, as it occurs for the galaxy distribution. In this case,
precise relations exist to connect the moments of the continuous field,
$\mu_n$, to
those, $m_n$, for the counts of the discrete realization. Under the minimal
assumptions of Poissonian sampling of
the continuous field, the discrete nature of the point distribution is
accounted for by the usual change of variable $\phi \to e^\phi -1$ in the
functional dependence of the generators $M(\phi)$ and $K(\phi)$
({  e.g.}, Layzer 1956; White 1979). As a consequence, new terms appear
when differentiating them, and eqs.(\ref{eq:kmu}) become
\ba\label{corr1}
\kappa_2 &=& m_2-1/\hat N\qquad {\rm ~(variance)}\,; \nonumber \\
\kappa_3 &=& m_3-3m_2/\hat N-1/\hat N^2 {\rm ~(skewness)}\,;\nonumber \\
\kappa_4 &=& m_4-6m_3/\hat N-3m_2^2+11m_2/\hat N-6/\hat N^3
{\rm ~(kurtosis)}
\ea
(Peebles 1980).
Note, however, that the Poissonian representation to account for
discreteness effects does not always provide a good prescription (Coles \&
Frenk 1991; Borgani \etal 1993). This is the case, for instance, when the
average count within $V_R$ is much less than unity, so that no informations
about the underlying continuous statistics can be recovered. A further
example is represented by the distribution of density peaks, which is far
from being a Poissonian sampling of the background continuous field.

In our analysis we take the window to be a cubic cell of size $R$. In this
case, if we make a partition of the simulation box into $N_b$ cubic cells,
the corresponding moments of counts reads
\be
m_n~=~{1\over N_b}\,\sum_{i=1}^{N_b}{(N_i-\hat N)^n
\over \hat N^n}\,,
\label{eq:mncc}
\ee
where $\hat N$ is the average cell count. Other authors considered
different windows. Davis \& Peebles (1983) analyzed the variance for the
count of CfA galaxies inside ``sharp" spheres, while Saunders \etal (1991)
evaluated the variance and the skewness for the QDOT IRAS sample for counts
within Gaussian spheres. In the following we will compare our analysis to
analogous results from QDOT galaxies. Therefore, we need a suitable
prescription to identify the size of the cubic cell to be associated to
that of the Gaussian sphere. As for the variance, its value evaluated
inside a Gaussian sphere of radius $R_{gs}$ coincide with that evaluated in
a cubic cell of side
\be
R_{cc}~=~(J_{2,cc}/J_{2,gs})^{1/\gamma}R_{gs}~=~3.35\,R_{gs}~~~~~~;~~~~~~
\gamma=1.8\,,
\label{eq:r2}
\ee
with obvious meaning of the indices, quite
independently of $\gam$. In general one expect a different relation
for the scales connecting the skewness estimate within the two
different windows, but we find numerically a relation very close
to (\ref{eq:r2})
within the MonteCarlo quadrature accuracy, again with a negligible
dependence on $\gamma$,
\be
R_{cc}~=~(J_{3,cc}/J_{3,gs})^{1/2\gamma}R_{gs}~=~3.25\,R_{gs}~~~~~~;~~~~~~
\gamma=1.8\,,
\label{eq:r3}
\ee
Note that, although the two prescriptions to identify the window sizes to be
associated to a given moment value do depend on the moment order itself, in
the case of interest we find only a rather weak dependence.

We focus now on the results for a representative subset of
the models, which
are in fair agreement with the angular APM correlation function. The three
fiducial models are F.14.3, 8.38.1, and 5.66.1 (see Table I for the
parameters relevant for these models).

\subsection*{\normalsize 3.2 ~~Results}
In Figure 3 we plot the scale-dependence of the variance for the three
bubble models considered, as compared with observational data. The error
bars for our models are not reported here for clarity; they have been
calculated as the ensemble scatter over ten different realizations, and are
plotted on the following figures. Open circles are for our models, while
filled circles are for the Stromlo-APM galaxies (Loveday \etal 1992)
and filled triangles for QDOT galaxies as analyzed by Efstathiou \etal
(1990; hereinafter QDOT90). The agreement on large scales
is fairly good, while a residual
discrepancy is always detected at small scales. The reason for this could
be that the variance for the APM galaxies has been evaluated in redshift
space and then corrected to real space according to the relation
$\sigma^2_{redsh.~sp.}=f(\Omega,b)\,\sigma^2_{real~sp.}$, where the
correction factor
\be
f(\Omega,b)~=~1+{2\over 3}\,{\Omega^{0.6}\over b}+{1\over 5}\,\left(
{\Omega^{0.6}\over b}\right)^2
\label{eq:fomb}
\ee
holds at the scales of linear clustering and gives an equal amount of
clustering amplification at all the scales (Kaiser 1987).
In eq.(\ref{eq:fomb}), $b$ is the linear biassing factor, which describe
the difference between relative fluctuations of galaxies and underlying
dark matter, $\delta_{gal}=b\delta_{DM}$.
Loveday \etal (1992) corrected their variance estimate by taking
$f(\Omega,b)=1.4$ at all the scales. However, this could
underestimate the clustering at small scales, where, instead, virial
motions are expected to suppress the redshift-space correlation ({
e.g.}, Fisher \etal 1993). It is however not clear how large this effect
should be in our case, since the bubble simulations does not provide galaxy
peculiar velocities. In any case, we do not believe that this small-scale
discrepancy represents a serious problem. A better agreement could be
reached by allowing for  a further 10\% of unclustered component, which
should decrease the $\sigma^2$ values, without significantly changing the
angular correlation (see Figure 2). However, here we are not interested in
finding the model which best fits the observational data, rather we are
analysing the ability of bubble geometries to reproduce the overall
features of the galaxy distribution.

In Figure 4 we report the relations of skewness $\mu_3$ and kurtosis
$\mu_4$ versus the variance $\sigma^2$. The models are compared with the
corresponding relations found in the analysis of Gatza\~naga (1992) for CfA
and SSRS galaxies (dashed lines) and of Bouchet, Davis
\& Strauss (1992) for the
Strauss \etal (1990) IRAS 1.2Jy sample (dot-dashed lines,
hereinafter IRAS 1.2). The scales analyzed
in the two observed samples are limited to about 30 $\hm$ for Gazta\~naga
(1992) and to about 70 $\hm$  for IRAS 1.2.
We also plot the
skewness-variance relation found by Saunders \etal (1991) in the QDOT
survey with Gaussian windows (hereinafter QDOT91).
The solid lines are the weighted best fit to
our model's values, considering only the points for which $\sigma^2<1$. The
fitting parameters are reported in Table I; considering also the points
with $\sigma^2>1$ would result in values still inside the errors. We
immediately note that, although the shallower power-law model 5.66.1
matches  the APM angular correlation function, it is unable to follow the
hierarchical relation (\ref{eq:hie}) in the fully linear regime.

Despite the apparently fair fit provided by the hierarchical relations,
Figure 5 shows that we cannot exclude deviation from hierarchicity, due to
the large error bars in both our model's and observational data. It appears
indeed that the plots of skewness and kurtosis as functions of the variance
may conceal even substantial departures from hierarchical scaling,
departures which, on the contrary, are best evidenced by plotting the
coefficients $S_3=\kappa_3/ \sigma^4$ and $S_4=\kappa_4/\sigma^6$ against
the box scale. Again open circles are for our simulations while filled
triangles are now for the QDOT91 analysis by Saunders \etal (1991), with
errors propagated from those originally quoted. Dashed lines and dot-dashed
lines denote the surprisingly small 2$\sigma$ uncertainty level for the
analysis of Gatza\~naga (1992) and the 1$\sigma$ level for IRAS 1.2,
respectively. We confirm with greater evidence that a shallow spectrum of
bubble radii does not provide hierarchical correlations at large scales,
and is totally at variance with the result at largest scale probed in
QDOT91. On the contrary, steep power-law models and fixed radius models
show a relative independence of $S_3$ with the scale, at least within the
statistical uncertainties, while the coefficient $S_4$ remains completely
undetermined at scales above $30\hm$. This does not allow a further
discrimination among the models.

Finally, in Figure 6 we display the results for ten simulations of the
models 8.38.1 and F.14.3 in the skewness-variance plane. These are compared
to the result from the QDOT91 analysis of Saunders \etal (1991) with the
largest Gaussian sphere, $R_{gs}=20 \hm$. Adopting the relation $R_{cc}=3.3
R_{gs}$ one gets $R_{cc}=66 \hm$, but since this scale identification
depends on a model for $\xi_2(r)$ which, at these scales, is quite
questionable, we plot the results for the ten simulations taking box sizes
of 60 $\hm$ (the rightward points) and 70 $\hm$ (the leftward points).  The
satisfactory agreement once again stresses the ability of certain bubble
models to account for the large scale clustering of galaxies. Notice,
however, that the results in QDOT91 have not been corrected
for the redshift distortion; applying a correction factor $f(\Omega,b)> 1$
would push down the variance by the same factor and the skewness by
$f(\Omega,b)^2$. Then the best agreement with our models would require
somewhat less power on large scales, to be obtained, for instance, by
increasing the unclustered component.

\section{\normalsize\bf Discussion and conclusions}
In this paper we realized a detailed clustering analysis of simulations of
bubbly galaxy distributions. Due to the purely geometrical character of our
simulations, we do not include small-scale gravitational clustering.
Rather, we concentrate on large ($>5\hm$) scales, where the clustering
should be in the linear regime or weakly non-linear regime, so that it can
be modelled by the geometry of the bubble distribution. In turn, the bubbly
geometry can be thought to be either the result of a primordial phase
transition or of quasi-linear gravitational dynamics, acting at
intermediate and large scales. The advantage of this kind of approach is
that we are able to generate very large simulation boxes of $450\hm$ aside,
which allow for a careful clustering analysis at large scales.

As a first test, we generated artificial Lick maps, by projecting galaxies
contained in two levels of replicated boxes (see CMPLMM and Moscardini
\etal 1993). The resulting Lick simulations include galaxies up to a
distance of $675\hm$ and is limited to the range of galactic latitude
$|b^{\mbox{\tiny II}}|\ge 45^\circ$. The possibility of reproducing the
observational setup of this angular sample enables us to realize a
comparison between the bubbly large-scale clustering and that of
observational data sets. As a remarkable result, by comparing the resulting
angular two-point correlation function, $w(\vartheta)$, with that of the
APM sample (Maddox \etal 1990), we find that its large-scale behaviour can
be fairly reproduced by several models.
Models with fixed radius, which are in any case less motivated, are found
to have troubles. In fact, choosing $R_M=12\hm$, the resulting
$w(\vartheta)$ amplitude agrees at $\vartheta \simeq 2^\circ$ with APM
data, but fails at larger scales. Vice versa, for larger radii the
agreement at $\vartheta \magcir 5^\circ$ is sensibly improved, but it turns
into an excess of clustering at smaller angles. Taking variable bubble
radii introduces a modulation of the correlation amplitude, which succeeds
at reproducing the observed $w(\vartheta)$ at all the relevant scales.
However, we are not able to discriminate at this level between steep and
shallow spectrum of radii.

In order to further discriminate between different models, we also applied
the statistics of moments of count in cells and compared the results with
similar analyses on observational data sets (Saunders \etal 1991,
Bouchet, Davis \& Strauss 1992,
Gatza\~naga 1992, Loveday \etal 1992). As for the scale-dependence of the
variance, we find that all the models, which satisfy the angular
correlation test, are also able to account for the Stromlo-APM redshift
data (Loveday \etal 1992). This is not surprising, since both angular and
spatial analyses of second order correlations are based on the same parent
sample. More interesting results appear as we consider at higher
correlation orders the scaling of skewness and kurtosis. In fact, the
bubble geometry turns out to develop a rather well defined hierarchical
scaling of higher order cumulants up to scales $\mincir 70\hm$,
with the exception represented by the large-scale behaviour of the shallower
spectra 4.78.1 and 5.66.1 (see Figures 4 and 5 and Table I). This result may
appear rather surprising, since hierarchical scaling is expected to have a
dynamical origin, both in the mildly and in the strongly non-linear regime
of gravitational clustering, while no dynamics is present in our
simulations.

It is however possible to show that hierarchical scaling is a necessary
consequence for any non-Gaussian distribution (like that implied by a
bubbly geometry) in the weak coherence regime. To show this, let us assume
to be at the scale $r$ in a regime of weak clustering for the fluctuation
field $\delta_r(\bx)$. Let us also consider the coarse-grained field
$\delta_R(\bx)$, smoothed at the scale $R\gg r$,
that can be considered as the linear superposition of many
independent processes $\delta_r(\bx)$:
\be
\delta_R(\bx)~=~{1\over p}\,\sum_{i=1}^p\delta_r(\bx_i)\,.
\label{eq:drd}
\ee
The corresponding moment of order $n$ is
\be
\mu_n(R)~=~\left<\delta_R^n(\bx)\right>~=~
{\left< \left[\sum_{i=1}^p\delta_r(\bx_i)\right]^n\right>
\over p^n}\,.
\label{eq:mnwc}
\ee
Since $\delta_r$ values are assumed to be nearly
independent, the average of all the cross products in the multinomial
expansion of eq.(\ref{eq:mnwc}) becomes negligible, so that
$\mu_n(R)=p^{-(n-1)}\mu_n(r)$. Therefore the same relation also holds between
the cumulants and, since $\sigma^2(R)\equiv \mu_2(R)=\mu_2(r)/p$,
it implies  the hierarchical scaling
\be
\kappa_n(R)~=~S_n\,[\sigma(R)]^{2(n-1)}\,.
\label{eq:hiwc}
\ee
with coefficients $S_n=\kappa_n(r)/[\sigma(r)]^{2(n-1)}$ fixed by the
statistics at the reference scale $r$ and independent of $R$.

It is clear that this argument to explain the presence of hierarchical
scaling crucially depends on the assumption of weak coherence. In most
cases, this simply means that the clustering is well in the linear regime,
$\sigma_R^2\ll 1$. This remarkably agrees with the results presented in
Table I for models having different percentages of unclustered galaxies. In
fact, as the unclustered component is increased, so to decrease the
correlation amplitude, the reliability of the hierarchical scaling improves
(see the values of $\eta_3$ and $\eta_4$ for the three models with
$p=4.5$). However, if the field $\delta(\bx)$ keeps coherent over large
scales, {  e.g.} due to the presence of intrinsic phase correlations,
coherent structures can still be present even in the small variance regime
and the hierarchical scaling does not show up. This is the case, for
instance, when large-scale phase correlation is imprinted in the initial
conditions or when dealing with primordial spectra providing a lot of
large-scale power, which rapidly develop extended structures in the
clustering pattern. This may explains, for instance,  why in the analysis
of N-body simulations by Bouchet \& Hernquist (1992) it is found that,
while the hierarchical relation is closely followed by the CDM model in the
mildly non-linear regime, the evidence of hierarchical behaviour on large
scales for the HDM model is much less clear. This is also the case for our
4.78.1 and 5.66.1 models, which do not develop hierarchical behaviours at
large scales. In fact, taking a shallower spectrum of bubble radii allows
the presence of very large bubbles, which generates a high degree of large
scale coherence.

In the introduction of this paper we addressed two questions concerning
whether clustering models based on bubble geometry are adequate to account
for the observed large-scale texture of the Universe and whether the
resulting clustering measures are sensitive to the choice of the model
parameters. Based on the results of our analysis we can give an affirmative
answer to both questions. It is however clear that our findings demand for
a dynamical interpretation about the origin of a bubbly galaxy
distribution. In particular one may ask whether they are imprinted as
initial conditions in the primordial fluctuations or develop through
gravitational evolution. In this respect, one could be tempted to suggest
that, while bubble generation through primordial phase transitions should
give rise to a narrow spectrum of radii (which is also
more likely to pass the cosmic microwave background constraints),
gravitational evolution of a
rather flat fluctuation spectrum should produce nearly spherical voids of
sizes comprised in a large range. It is however clear that, before rush to
definite conclusions on this point one should wait, from the observational
side, for more extended data sets, which were able to encompass the largest
scale involved by coherent structures, and, from the theoretical side, for
more rigorous dynamical descriptions like those provided by running large
N-body simulations.

\vspace{1.truecm}
\noindent
{\bf Acknowledgements.} We wish to thank Dr. Steve Maddox for providing us
with the digitized version of the APM correlation data.

\newpage
\centerline{\bf References}

\frenchspacing
\noindent
Achilli, S., Occhionero, F., Scaramella, R., 1985, \apj, 299, 577\\
Amendola, L., Occhionero, F., 1993, \apj, in press\\
Babul, A., White, S.D.M., 1991, \mnras, 253, 31{\tiny P}\\
Bertschinger, E., Dekel, A., Faber, S.M., Dressler, A.,
Burstein, D., 1990, \apj, 364, 370 \\
Blumenthal, G.R., Faber, S.M., Primack, J.R., Rees, M.J., 1984, \nat,
311, 517 \\
Borgani, S., Coles, P., Moscardini, L., Plionis, M. 1993, \mnras,
submitted\\
Bouchet, F., Davis, M., Strauss, M., 1992, in Proc. of the 2nd DAEC Meeting on
the Distribution of Matter in the Universe, eds. G.A. Mamon, \& D. Gerbal,
p.287 (IRAS 1.2)\\
Bouchet, F. R. Hernquist, L., 1992, \apj 25, 400\\
Bower, R.G., Coles P., Frenk C.S., White, S.D.M., 1993, \apj, 405, 403\\
Cen, R., Gnedin, N.Y., Kofman, L.A., Ostriker, J.P., 1992, \apj, 399, L11\\
Coles, P., Frenk, C.S., 1991, \mnras, 253, 727\\
Coles, P., Moscardini, L., Plionis, M., Lucchin, F., Matarrese, S.,
Messina, A., 1993, \mnras, 260, 572 (CMPLMM)\\
Collins, C.A., Heydon-Dumbleton, N.H., MacGillivary, H.T., 1989,\mnras,
236, 7{\tiny P} \\
Couchman, H.M.P., Carlberg, R.G., 1992, \apj, 389, 453 \\
Davis, M., Efstathiou, G., Frenk, C.S., White, S.D.M., 1992, \nat, 356,
489\\
Davis, M.,Meiksin, A., Strauss, M., Nicolaci da Costa, L., Yahil, A., 1988,
\apj, 333, L9\\
Davis, M., Peebles, P.J.E., 1983, \apj, 267, 465\\
Efstathiou, G., Ellis, R.S., Peterson, B.A., 1988, \mnras, 232, 431\\
Efstathiou, G., Kaiser, N., Saunders, W., Lawrence, A., Rowan-Robinson,
M., Ellis R. S., Frenk C. S., 1990, \mnras, 247, 10{\tiny P} (QDOT90)\\
Efstathiou, G., Sutherland, W.J., Maddox, S.J., 1990, \nat, 348, 705\\
Feldman, H., Kaiser, N., Peacock, J., 1993, preprint\\
Fisher, K. B., Davis, M., Strauss, M. A., Yahil, A., Huchra,
J.P., 1993, \apj, 402, 42\\
Fry, J.N., 1984a, \apj, 277, L5\\
Fry, J.N., 1984b, \apj, 279, 499\\
Gazta\~naga, E., 1992, \apj, 398, L17\\
Hamilton, A.J.S., 1988, \apj, 332, 67\\
Ikeuchi, S., 1981, Publ. Astr. Soc. Japan, 33, 211\\
Juszkiewicz, R., Bouchet, F.R., 1992, in Proc. of the 2nd DAEC Meeting on
the Distribution of Matter in the Universe, eds. G.A. Mamon, \& D. Gerbal,
p.301\\
Kaiser, N., 1987, \mnras, 227, 1\\
Klypin, A., Holtzman, J., Primack, J., Reg\"os, E., 1992, preprint SCIPP
92/52\\
Kolb, E.W., 1991, Phys. Scr., T36, 199\\
La, D., 1991, \pl, B265, 232\\
La, D. Steinhardt, P.J., 1989, \prl, 62, 376\\
Lahav, O., Itoh, M., Inagaki, S., Suto, Y., 1993, \apj, 402, 387\\
Layzer, D., 1956, \apj, 61, 383\\
Liddle, A., Lyth, D.H., 1993, Phys. Rep., in press \\
Liddle, A. R., Wands D., 1991, \mnras, 253, 637\\
--- 1992, preprint SUSSEX-AST-92/2-2\\
Loveday, J., Efstathiou, G., Peterson, B. A., Maddox,
S.J., 1992, \apj, 400, L43\\
Lucchin, F., Matarrese, S., 1985, \pr, D32, 1316\\
Maddox, S.J., Efstathiou, G., Sutherland, W.J., Loveday, J. 1990,
\mnras, 242, 43{\tiny P} \\
Moscardini, L., Borgani, S., Coles, P., Lucchin, F., Matarrese, S.,
Messina, A., 1993, \apj, submitted\\
Moscardini, L., Matarrese, S., Lucchin, F., Messina, A., 1991, \mnras,
248, 424\\
Ostriker, J.P., Cowie, L.L., 1981, \apj, 243, L127\\
Peacock, J.A., 1991, \mnras, 253, 1{\tiny P}\\
Peebles, P.J.E., 1980, The Large Scale Structure of the Universe.
Princeton University Press, Princeton\\
Plionis, M., Barrow, J.D., Frenk, C., 1991,\mnras, 249, 662 \\
Saunders, W., Frenk, C., Rowan-Robinson, M., Efstathiou,
G., Lawrence, A., Kaiser, N., Ellis, R., Crawford, J., Xia, X.-Y., Parry,
I., 1991, \nat, 349, 32 (QDOT91)\\
Strauss, M.A., Davis, M., Yahil, A., Huchra, J.P., 1990, \apj, 361, 49 \\
Valdarnini, R., Bonometto, S.A., 1985, \aaa, 146, 235\\
Van de Weygaert, R., Icke, V., 1989, \aaa, 213,1\\
Van de Weygaert, R., 1993, preprint\\
Vittorio, N., Juszkiewicz, R., Davis, M., 1986, \nat, 323, 132 \\
Weinberg, D.H., Cole, S., 1991, \mnras, 259, 652\\
Weinberg, D.H., Ostriker, J. P., Dekel, A., 1989, \apj, 336, 9\\
White, S.D.M., 1979, \mnras, 186, 145\\
White, S.D.M., Efstathiou, G., Frenk, C.S., 1993, preprint\\
White, S.D.M., Frenk, C.S., Davis, M., Efstathiou, G., 1987, \apj,
313, 505 \\
Wright, E.L., \etal, 1993, preprint\\
Yoshioka, S., Ikeuchi, S., 1989, \apj, 341, 16\\

\newpage
\centerline{\bf Figure captions}

\vspace{0.3truecm}
\noindent
{\bf Figure 1.} Slices of the bubbly galaxy distribution for different models
(see Table I for the parameters of each model). Each slice is $15\hm$ thick
and all the realizations correspond to the same choice of initial
phase-assignment.

\vspace{0.3truecm}
\noindent
{\bf Figure 2.} The angular correlation function, $w(\vartheta)$, for
simulations of Lick map, obtained by projecting the three-dimensional
bubble simulations. Open circles refer to the APM correlation, rescaled to
the depth of the Lick map, as provided by Maddox \etal (1990). Filled
symbols are for our simulations. They are plotted only for
$\vartheta \ge 1.5^\circ$, which correspond to
the physical scales of linear clustering at the depth of the Lick map.

\vspace{0.3truecm}
\noindent
{\bf Figure 3.} The scale dependence of the variance for counts in cubic
cells. Open circles are for the bubble simulations, which are obtained
after averaging over ten realizations. Filled triangles are
for the IRAS QDOT analysis by Efstathiou \etal (1990) and filled circles
are for Stromlo-APM galaxies (Loveday \etal 1992). For reason of clarity,
we do not plot errorbars for our data.

\vspace{0.3truecm}
\noindent
{\bf Figure 4.} The variance-skewness (left panels) and the kurtosis-skewness
(right panels) relations. Open circles refer to our simulations, with
errorbars denoting the r.m.s. scatter over an ensemble of ten realizations.
Heavy bars for the skewness are the QDOT data by Saunders \etal (1991). The
dot-dashed lines are the best fits to the analysis of the
IRAS 1.2Jy survey by Bouchet, Davis \& Strauss
(1992) and dashed lines refer to the analysis of CfA and SSRS samples by
Gatza\~naga (1992). The solid lines are the best fits to our data, according
to the parameters reported in Table I.

\vspace{0.3truecm}
\noindent
{\bf Figure 5.} The scale dependence of the $S_3$ (left panels) and $S_4$
(right panels) hierarchical coefficients of eq.(\ref{eq:hie}). Open
circles are for our models, while dot-dashed and dashed lines delimit
the uncertainty bands in the IRAS 1.2Jy analysis by
Bouchet, Davis \& Strauss (1992) and
Gatza\~naga (1992), respectively. Filled triangles for $S_3$ refer to the
Saunders \etal (1991) analysis of QDOT data.

\vspace{0.3truecm}
\noindent
{\bf Figure 6.} The variance-skewness relation for the QDOT data at the
largest considered scale, corresponding to the size of $66\hm$ for a cubic
box. Each open circle refers to the result for one realization of the
corresponding bubble model. Leftmost circles are for a box-size of $70\hm$,
while rightmost ones are for the $60\hm$ scale.

\newpage

\baselineskip 30pt
\large
\begin{tabular}{lllllllll}
\multicolumn{9}{c}{\bf Table I }\\ ~\\
 model &$~p~$ 	& $R_M$	& $R_m$ & $~f_u~$ &   $S_3~(\Delta S_3)$ & $~\eta_3~
(\Delta \eta_3)$ & $S_4~(\Delta S_4)$ &$~\eta_4~(\Delta \eta_4)$\\ ~\\
4.78.1  &4.5	& 78	 & 5.5	 &  0.1	 & 1.44(40) & 2.47(46)&1.41(1.03)&
3.26(2.62)\\
4.78.2  &4.5	& 78	 & 5.5	 &  0.2	 & 1.64(22) & 2.32(32)&1.89(65)&
3.05(1.21)\\
4.78.3  &4.5	& 78	 & 5.5	 &  0.3	 & 1.85(27) & 2.22(25)&2.37(87)&
3.00(1.00)\\ ~\\
5.66.1& 5 & 66	& 6.5  &0.1 & 1.84(1.10)& 2.42(70)& 2.25(4.04)& 3.21(2.72)\\
{}~\\
8.36.1 & 8 & 36	& 9. &0.1 &1.77(55)	&1.97(33) &  3.43(3.29)
&2.81(1.15)	\\
8.38.1 & 8 & 38	& 9.5 &0.1 &1.77(42)	&1.97(31) &  3.41(2.14)
&2.81(1.09)	\\ ~\\
F.12.3 & -- & 12	& --  & 0.3 &  2.11(30)  & 1.85(19)&4.70(1.10)	&
2.53(51)\\
F.14.3 & -- & 14	& --  & 0.3 &  2.20(60)  & 1.85(33)&5.23(3.56)	&
2.54(88)\\
\end{tabular}

\end{document}